\documentstyle[12pt]{article}
\begin{document}
\centerline{\bf S.P.Novikov}
\vspace{5mm}
\centerline{\bf Various Doublings of  Hopf Algebras}
\centerline{\bf Algebras of Operators on  Quantum Groups and
 Complex Cobordisms}
\vspace{7mm}
Let us consider the pair of the dual
Hopf algebras $X$, $X^{\ast}$ over
 the commutative associatiative
ring $k$ with $1\in k$
 (or field)
with the diagonals (comultiplications)
 $\Delta$, antipods $s$ and augmentation
 $\epsilon$. Let the algebra $M$ be
 an $X$-module compatible with $\Delta$:
$$x(uv)=\sum x^1_i(u)x^{11}_i(v)$$
$$\Delta(x)=\sum x^1_i\bigotimes x^{11}_i$$

We call such module $M$ a {\bf Milnor module}
(who discovered from that the Hopf
 property of the Steenrod algebra about
 1957 [1]). It was a starting point of
 the algebraic theory
 of the Hopf algebras. They were introduced
 by A.Borel in 1954 as a result of the
algebraic analysis of the Hopf theorem
 on the cohomology rings of H-spaces,
which he extended to the finite fields.

We introduce the operator algebras
 generated by the left multiplications
 $u$, right multiplications $v$ and
 by the action of $X$: $A=(M\bigotimes
M^1)X$ (left modules), $B=X(M^1\bigotimes
M)$ (right modules):
$$m\rightarrow ux(m)v$$
with the commuting relations:
$$x(u\bigotimes v)=\sum(x^1_q(u)\bigotimes
x^{111}_q(v))x^{11}_q$$
(left modules)
$$(u\bigotimes v)x=\sum x^{11}_q
(x^1_q(u)\bigotimes x^{111}_q(v))$$
(right modules).

Here $\Delta^2=\sum x^1_q\bigotimes
 x^{11}_q\bigotimes x^{111}_q$,
 $M^1$ is the ''inverted algebra''
with the multiplication $u\cdot v=vu$.

These algebras are associative if $X$,
$X^{\ast}$, $M$ are. We always suppose that.
The subalgebras $A_L=(M\bigotimes 1)X$,
$B_R=X(1\bigotimes M)$ will be considered here.
We call them the {\bf O-doubles}.

{\bf Basic examples}: let $M=X^{\ast}$.
 We use the general notations for the right
and left multiplication operators
 $L_a(b)=R_b(a)=ab$. Let $L^{\ast}_x$,
 $R^{\ast}_x$ be the adjoint operators
 of $X$ on the space $X^{\ast}$.

{\bf Lemma 1}. $M=X^{\ast}$ is the left
 (corr. right)
Milnor module for the representations
$R^{\ast}_{s^{2k}x}$,
 $L^{\ast}_{s^{2k}x}$ of the Hopf
algebra $X$; it is the left (corr. right)
 Milnor module of the Hopf algebra
$X^t$ for the representations
$L^{\ast}_{s^{2k+1}x}$,
 $R^{\ast}_{s^{2k+1}x}$. Here $X^t$ is the
same algebra with the ''inverted diagonal''
  $\Delta^t=\sum x^{11}_i\bigotimes
x^{1}_i$ (it might be only bialgebra
  if  $s^{-1}$ does not exist).

 We have $X^{t\ast}=
X^{\ast 1}$. The representations
 $L^{\ast}$ and $R^{\ast}$
commutes with each other.
Therefore the algebra $X^{\ast}=M$ is the
left Milnor module over the Hopf algebra
 $X\bigotimes X^t$.

In the full left operator algebra $A_L=(X^{\ast}
\bigotimes 1)(X\bigotimes X^t)$ there is
 an important subalgebra $C=(X\bigotimes 1)
\Delta(X)$ with the relations
$$xu=\sum [\rho_{1x^1_q}\rho_{2x^{111}_q}(u)]x^{11}_q$$

Here $\rho_1$ and $\rho_2$ are the commuting
 representations
 of $X,X^t$. They determine the
 {\bf ''ad-modules''} over the algebra $X$
 on the diagonal $\Delta(X)\subset
X\bigotimes X^t$:
$$\rho_x(u)=\sum  \rho_{1x^1_i}\rho_{2x^{11}_i}(u)$$

such that
$$\rho_x(uv)=\sum [\rho_{1x^1_q}\rho_{2x^{111}_q}(u)]
\rho_{x^{11}_q}(v)$$

For the cocommutative $X$  ad-modules are
 the Milnor modules; for $M=X^{\ast}$,
$\rho_1=R^{\ast}_x$, $\rho_2=L^{\ast}_{sx}$
 the algebra $C$ above is exactly the
 {\bf Drinfelds Quantum Double}
[1,2,3], which is a Hopf algebra with the
comultiplication
$\Delta(ux)=\Delta^t(u)\Delta(x)$.

{\bf O-Doubles} appear as the Operator algebras
 (differential operators on the Lie groups and
  difference operators on the discreet groups).

An important example for the ring $k=Z$
gives an algebra $A^U=\Lambda X$,
of all ''cohomological'' operations
 in the Complex Cobordism Theory.
It was computed by the author in 1966 [5].

Here $X$ is the so-called ''Landweber-Novikov''
 algebra found in [5,6], $\Lambda$ is the
 $U$- cobordism ring ''for the point''
found by Milnor and author about 1960.

The representation of the Hopf algebra $X$
 on the ring $\Lambda$
was found geometrically [5]. In fact we have
$\Lambda\subset X^{\ast}$, $\Lambda\bigotimes
Q=X^{\ast}\bigotimes Q$. The Hopf properties
 of the pair $\Lambda\subset X$ and the relation of
 the algebra $A^U$
to the differential operators on some infinite-
dimensional Lie group were discovered
in [7] by Bukhstaber and Shokurov.

{\bf Proposition}: The ''geometric''
 representation
 of the algebra $X$ on $\Lambda$ is
 exactly $R^{\ast}_x$
for the algebra $A^U$.

The present author found this recently
 and proved as a consequence from [7]
(it was not observed by the authors of [7]
but may be easily deduced from this paper).

O-doubles are not the Hopf algebras,
 but they have some ''almost Hopf''
properties for the representations
$\rho= L^{\ast},R^{\ast}$. For $A_L=
X^{\ast}X$ the adjoint space is
obviously $A^{\ast}=XX^{\ast}$. The
 multiplication $\Psi :A_L\bigotimes
A_L\rightarrow A_L$ generates the
 comultiplication $\Psi^{\ast} :
A^{\ast}\rightarrow A^{\ast}\bigotimes
A^{\ast}$.

{\bf Lemma 2}: For the representation $\rho
 =R^{\ast}_x$ the following formula is valid:
$$\Psi^{\ast} (xu)=\Delta (x)R\Delta (u)$$

Here $\Psi^{\ast} (1)=R=\sum e^i\bigotimes e_i$,
 $(e^i,e_j)=\delta^i_j$, $(u,x)=\epsilon R^{\ast}
_x(u)$- the canonical scalar product, $e_j$ is the
 basis of $X$.

{\bf Lemma 3}: For the representations $\rho_1=
R^{\ast}_{s^{2k}x}$ and $\rho_2=L^{\ast}_{s^{2k}u}$
 the corresponding O-doubles $A_L=X^{\ast}X=B_L$
 exactly coinside. The same for the representations
 $\rho_1=L^{\ast}_{s^{2k+1}x}$ and $\rho_2=
R^{\ast}_{s{2k+1}u}$: $A^1_L=X^{\ast 1}X^t=B^1_R$.

{\bf Lemma 4}: The following antihomomorphisms
 are well-defined
$$X^{\ast}X\rightarrow X^{\ast}X^t\rightarrow
X^{\ast}X$$

where $ux\rightarrow s^{2m+1}(x)s^{2l+1}(u)$,
 $\rho_1=R^{\ast}_{s^{2k}x}$, $\rho_2=L^{\ast}_
{s^{2n+1}x}$, $\rho_3=R^{\ast}_{s^{2k^1+1}x}$,
 $m+l+2=k-n$ for the first arrow and $k^1-n=
-(l+m)$ for the second arrow.

{\bf Theorem 1}: Let the antipod $s$ is invertible.

1. The antiisomorphisms are well-defined:

a) {\bf Formally adjoint operators}
$$A_L(=X^{\ast}X)\rightarrow A^1_L(=X^{\ast 1}X^t)
\rightarrow A_L$$
where $\rho_1=\rho_2=\rho_3=R^{\ast}_x$,
 $ux\rightarrow s^{-1}(x)u$ for the first arrow
 and $s$ should be replaced by $s^{-1}$ for the
second arrow.

b) {\bf Hermitian adjoint operators}
$$A_L(=X^{\ast}X)\rightarrow A^{+}_L(=X^{\ast}X^t)
\rightarrow A_L$$
where $\rho_1=\rho_3=R^{\ast}_x$, $\rho_2=
L^{\ast}_{s^{-1}x}$, $ux\rightarrow s^{-1}(x)s(u)$
 or $ux\rightarrow s(x)s^{-1}(u)$ for the both arrows.

2. The algebras $A^{+}_L=X^{\ast}X^t$,
 $\rho=L^{\ast}_{s^{-1}x}$ and $A^{\ast}=XX^{\ast}$,
$\rho=R^{\ast}_u$ are canonically isomorphic.

The proof uses the lemmas 3 and 4.

The lemma 3 shows also that the  right analog
 of the $p$-representation is the action
of the algebra $X^{\ast}X$ on $X^{\ast}$:

$u\rightarrow L_u$(left multiplication),
$x\rightarrow R^{\ast}_x$ (r-differentiation)

The right analog of the $x$-representation
is the action of the same algebra on $X$:

$x\rightarrow R_x$ (right mul;tiplication)
$u\rightarrow L^{\ast}_u$ (l-differentiation).
The multiplication and differentiation
are {\bf Fourier-dual}.
The element $R$ from the lemma 2 is
 equal to $exp[p\bigotimes x]$ for the
commutative Hopf algebra with the single
 primitive generator
$x$; $p$ is the dual primitive generator,
 $sx=-x,sp=-p$.

For the cocommutative Hopf algebras $X$
 the most important problems involving
 the ''almost Hopf'' properties of
 the O-doubles (for example, $A^U$ in
the complex cobordism theory) were
 connected with the ''Lie-semigroup''
containing the all multiplicative elements
 $a(uv)=a(u)a(v)$, where $a$ may belong
 to some formal complition of the O-double,
defined in such a way that the action
on the dual $X^{\ast}$ is well-defined
(the right $U$-analogs of the Adams
operations, Chern character and Projectors
$a^2=a$
-see [5,7,8]).

 The analogous problems
 are very interesting also for the
 almost cocommutative or quasitriangular
 ''Hopf-Drinfeld'' algebras (quantum
groups with well-defined abstract $R$-matrix
determining the commutation rules
 in the algebra $X^{\ast}$).

{\bf References}

1. Milnor J..Ann of Math, vol 90 (1958), pp 272-280

2. Drinfeld V.G.. Proc Int Cong Math (1986-Berkleley),
 vol 1 Amer Math Soc Publ (1987), pp 798-820

3. Drinfeld V.G. Algebra and Analysys
vol 1 iss 2 (1989) pp 30-46
(in Russian)

4. Madjid Sh.. Int Journ of Modern Phys A vol 5 n 1
 (1990) pp 1-91

5. Novikov S.P. Isv AN SSSR (ser math) vol 31 (1967)
 pp 855-951

6. Landweber P.. Trans Amer Math Soc vol 27 n 1 (1967)
 pp 94-110

7. Bukhstaber V., Shokurov S.. Funk Analysys Appl vol 12
 iss 3 (1978) pp 1-14

8. Quillen D.. Bull of Amer Math Soc vol 75 n 6 (1969)
 pp 1293-1298

{\vspace{7mm}}
{\bf Remark}: After the discussion with L.Faddeev
 I realized that
 some construction of the special O-double
 was found  independently
in the recent paper of Faddeev and Alexeev in the
special important examples (of course they
did not know the cobordism theory). Their paper does
not use the the Hopf-algebraic terminology at all
(the relations in that algebra
 were written by the complicated special formulas).

M.Semenov-Tyanshansky wrote now
the paper more closed to our
ideas but still he does not use
 some elementary algebraic definitions  replacing
them by the complicated formulas in the special
 examples (his paper will be published very soon
 in the Theor Math Phys, November 1992).

Yu. Manin wrote a parer ''Note on Quantum groups
 and Quantum De Rham complexes'', 
 MPI fur mathematics, Bonn 
 (1991), dedicated mainly to some analogs 
 of differential forms. There is a discussion 
 concerning the rings 
 of differential operators 
 and scew tensor products in his paper. Some of his ideas are
 parallel to ours, but the main direction is different.

 Our paper already
 appeared in The Uspechi Math Nauk (1992) iss 5
(September-October) in Russian.

\end{document}